\documentclass[pdflatex,sn-mathphys-num]{sn-jnl}
\usepackage{graphicx}
\usepackage{multirow}
\usepackage{amsmath,amssymb,amsfonts}
\usepackage{amsthm}
\usepackage{mathrsfs}
\usepackage[title]{appendix}
\usepackage{xcolor}
\usepackage{textcomp}
\usepackage{manyfoot}
\usepackage{booktabs}
\usepackage{algorithm}
\usepackage{algorithmicx}
\usepackage{algpseudocode}
\usepackage{listings}
\theoremstyle{thmstyleone}

\theoremstyle{thmstyletwo}

\theoremstyle{thmstylethree}

\begin{document}

\title[Article Title]{Thermodynamic Cost of Regeneration in a Quantum Stirling Cycle}

\author*[1]{\fnm{Ferdi} \sur{Altintas}}\email{ferdialtintas@ibu.edu.tr; ferdialtintas.mail@gmail.com}

\affil*[1]{\orgdiv{Department} of Physics, \orgname{Bolu Abant İzzet Baysal University}, \orgaddress{\city{Bolu}, \postcode{14030}, \country{Türkiye}}}

\abstract{We study the standard four-stroke regenerative quantum Stirling heat engine cycle, which assumes local thermal equilibrium at each stage, within the standard weak-coupling, Markovian open quantum system framework. We point out that the regeneration process is not thermodynamically free in a reduced open-system description, and we treat the required work input as an explicit regeneration cost by modifying the cycle efficiency accordingly. We consider two working substances—a single spin-$1/2$ and a pair of interacting spin-$1/2$ particles—and investigate the cycle performance by taking the regeneration cost at its minimum value set by the Carnot heat-pump limit. For comparison, we also analyze the conventional Stirling cycle without regeneration under the same conditions. The super-Carnot efficiencies reported under the cost-free regeneration assumption disappear once the regeneration cost is included: the modified efficiency stays below the Carnot bound, while still remaining higher than the efficiency of the conventional Stirling cycle. For the conventional Stirling cycle, we provide a rigorous Carnot bound using quantum relative entropy, whereas for the regenerative cycle we derive a sufficient lower bound on the regeneration cost that guarantees thermodynamic consistency. Finally, we suggest three candidate quantum regenerator models for future work.}


\pacs[PACS number(s)]{05.70.-a, 07.20.-n, 07.20.Pe, 44.90.+c}

\maketitle

\section{Introduction}\label{sec1}
Quantum thermodynamics~\cite{deffner2019}, as an important component of quantum technologies, has progressed markedly in recent decades, offering new insights into miniaturized quantum systems~\cite{myers2022,bhatta2021,cangemi2024} and their exploration as quantum thermal machines~\cite{quan2007,quan2009,huang2014,gupta2021,lin2003,phung2025,xia2024,yin2018,he2002,lin2003b,lin2004,lin2005,huang2002,yin2020,yin2025,yin2017,chen2002,wu1998,zhao2017,paula2025,alcala2026,wang2024,xu2025,purkait2022,soufy2025,cruz2023,cakmak2023,das2023,castorene2025,castorene2025-2,hamedani2021,aydiner2021,rastegar2025,scully2003,dillen2009,gardas2015,thomas2018,guff2019,abah2017,altintas2025,campisi2016,li2025,peterson2019,klaers2017,assis2019,martinez2016,camati2019,feldmann2006,pedram2023,kose2019}. In particular, it aims to understand how genuinely quantum features—such as coherence, superpositions of states, and quantum correlations—can modify thermal properties and performance at the nanoscale. Recent experimental progress has accelerated the realization of nanoscale thermal machines, motivating systematic tests of when quantum effects and engineered environments can provide advantages over classical designs~\cite{li2025,peterson2019,klaers2017,assis2019,martinez2016}.

Within this framework, the formulation of thermodynamic processes (e.g., isothermal, adiabatic, isochoric, and isobaric) and the resulting quantum thermal cycles—such as Carnot, Otto, Stirling, Brayton, and Diesel—have been extensively studied across a broad range of quantum platforms, ranging from ion traps and nuclear magnetic resonance to superconducting devices and photovoltaic systems~\cite{myers2022,bhatta2021,cangemi2024,quan2007,quan2009}. Among these, the Otto and Carnot cycles have attracted particular attention, having been studied extensively in theory~\cite{myers2022,bhatta2021,cangemi2024,quan2007,quan2009} and, in several instances, demonstrated experimentally~\cite{peterson2019,klaers2017,assis2019,martinez2016}. The influence of quantum features—either in the working substance or in the reservoirs—on performance has been investigated widely; in particular, quantum-reservoir models suggest that engineered non-equilibrium resources can yield apparent efficiencies exceeding the ideal Carnot value if their preparation and maintenance costs are neglected~\cite{cangemi2024,rastegar2025,scully2003,dillen2009,klaers2017}. When these thermodynamic costs are properly included, the overall efficiency remains bounded by the Carnot limit~\cite{cangemi2024,gardas2015,thomas2018}. In addition, a variety of quantum-control protocols have been proposed to enhance cycle performance, including quantum lubrication to mitigate friction-like irreversible losses~\cite{feldmann2006}, shortcuts to adiabaticity to emulate slow adiabatic parameter changes in finite time~\cite{guff2019,abah2017}, shortcuts to equilibration to accelerate the relaxation of open quantum systems toward equilibrium~\cite{pedram2023}, and heat-bath algorithmic cooling schemes to increase cycle speeds~\cite{kose2019}.

In this work, we study a regenerative quantum Stirling heat engine cycle, focusing on the operation of the regenerator and its associated thermodynamic cost. In classical Stirling engines, the regenerator is introduced as an efficiency-enhancing component and is typically treated as a reversible heat transfer device, i.e., a passive element through which the working gas flows without an overall increase in its total entropy~\cite{zohuri2018,cengel2019}. Under this idealized assumption, the regenerative Stirling cycle yields particularly notable results: it can offer a higher mean effective pressure than the Carnot cycle, and—in the ideal gas limit with a perfect, lossless regenerator—it can attain the ideal Carnot efficiency~\cite{zohuri2018,cengel2019}. Accordingly, numerous works have examined, across a wide range of quantum platforms, a quantum Stirling cycle in close analogy with its classical four-stroke counterpart, consisting of two quantum isothermal strokes and two regenerative quantum isochoric strokes~\cite{huang2014,gupta2021,lin2003,phung2025,xia2024,yin2018,he2002,lin2003b,lin2004,lin2005,huang2002,yin2020,yin2025,yin2017,chen2002,wu1998,zhao2017,paula2025,alcala2026}. Similarly to the classical interpretation, the regenerator is typically modeled as a passive heat buffer that stores heat during one isochoric stroke and releases it during the other, thereby improving cycle performance. Within this framework, quantum regenerative cycles can outperform their non-regenerative counterparts, and in certain scenarios their efficiencies have even been reported to surpass the ideal Carnot bound~\cite{huang2014,phung2025,zhao2017}.

In this study, we revisit the standard four-stroke regenerative quantum Stirling cycle, which assumes local thermal equilibrium at each stage~\cite{huang2014,gupta2021,lin2003,phung2025,xia2024,yin2018,he2002,lin2003b,lin2004,lin2005,huang2002,yin2020,yin2025,yin2017,chen2002,wu1998,zhao2017,paula2025,alcala2026}. Departing from the standard cost-free passive heat buffer assumption, we focus on the functioning of the regenerator and account for a finite thermodynamic cost associated with heat regeneration. By explicitly introducing this mandatory work input, we demonstrate that the previously reported super-Carnot efficiencies are merely artifacts of an incomplete thermodynamic accounting~\cite{huang2014,phung2025,zhao2017}. Once this hidden cost is integrated into the cycle, thermodynamic consistency is restored, and the efficiency remains strictly bounded by the Carnot limit. Our motivation is as follows: In the four-stroke formulation of a regenerative quantum cycle, it is typically assumed that, at the end of each regenerative isochoric stroke, the working substance—after interacting with the regenerator—thermalizes to the temperature of the subsequent isothermal bath (i.e., the cold bath at $T_c$ after the first isochore and the hot bath at $T_h$ after the second), so that its reduced state can be represented by a Gibbs (thermal) state~\cite{myers2022,huang2014}. Within the standard weak coupling, Markovian open quantum system framework, this approach effectively models the regenerator as a high heat capacity reservoir with an effective temperature, i.e., as an effective heat bath~\cite{dann2018,breuer2002}. Consequently, the regenerator plays the role of a cold bath during the first regenerative isochoric stroke and that of a hot bath during the second. For this reason, regeneration requires that heat stored at a low temperature level be made usable at a higher temperature level. In other words, upgrading energy/heat available at the $T_c$ level to the $T_h$ level is, from a second-law perspective, a heat-pumping (heat-upgrading) problem: it cannot occur spontaneously and therefore must be driven by an external source of work.

Motivated by recent thermodynamic cost accounting approaches~\cite{cangemi2024,gardas2015,thomas2018,guff2019,abah2017}, we account for the hidden work input as an explicit regeneration cost within the working substance thermodynamic description, and we use it to redefine the cycle efficiency accordingly. By taking the regeneration cost at its minimum value set by the Carnot heat pump limit~\cite{cengel2019}, we revisit the efficiency of the regenerative Stirling cycle for two working substances: a single spin-$1/2$ and a pair of interacting spin-$1/2$ particles. As a benchmark, we consider the conventional Stirling cycle without regeneration and evaluate its efficiency under identical conditions, thereby determining whether the regenerator retains a performance advantage once the regeneration cost is included.

Our results show that, when the regeneration cost is neglected, the thermodynamic efficiency can exceed the ideal Carnot bound; however, once the cost term is included, the resulting modified efficiency is reduced and remains below the ideal Carnot efficiency, while still exceeding the efficiency of the conventional (non-regenerative) Stirling cycle. While previous numerical studies of conventional quantum Stirling cycles across various working media have found that the efficiency is always bounded by the ideal Carnot bound~\cite{alcala2026,purkait2022,wang2024,xu2025,soufy2025,cruz2023,cakmak2023,das2023,castorene2025,castorene2025-2,hamedani2021,aydiner2021,rastegar2025}, in this work we provide an analytical proof—using a quantum relative entropy formulation—that the ideal Carnot efficiency constitutes a rigorous upper bound. In contrast, for the regenerative Stirling cycle we derive an explicit sufficient lower bound on the regeneration cost that guarantees the cycle efficiency always remains below the ideal Carnot bound. Finally, to motivate future work, we propose three possible quantum models in which the regenerator is a proper quantum system and is treated as an active component of the cycle, responsible for implementing the isochoric strokes and regenerating heat—based on a non-Markovian heat reservoir, an auxiliary quantum system, and a collisional model—and we briefly discuss the physical motivation underlying each proposal.

\section{Regenerative Quantum Stirling Cycle}\label{sec2}

\begin{figure}[h]
\centering
\includegraphics[width=1.0\textwidth]{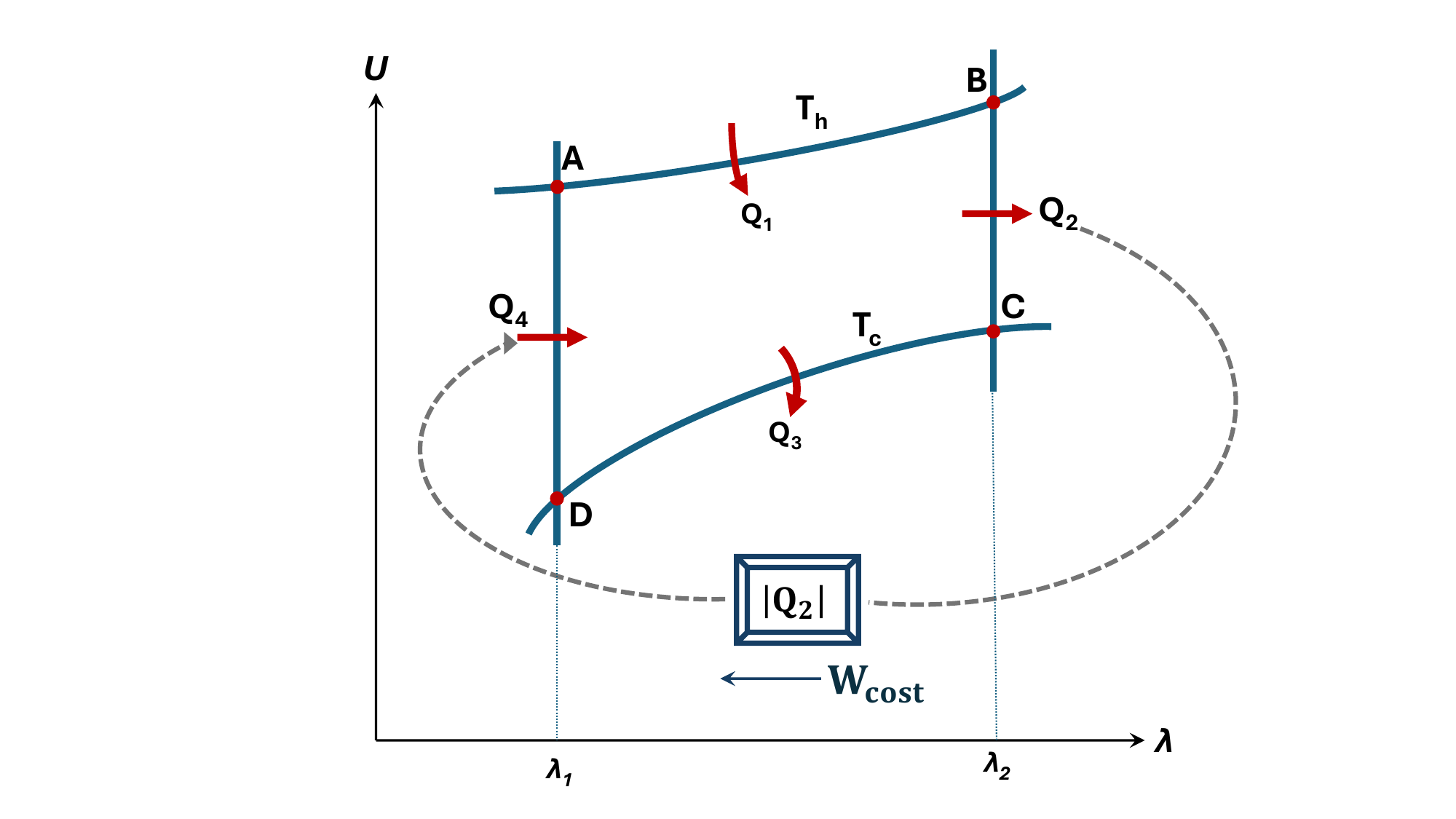}
\caption{The schematic representation of the regenerative quantum Stirling cycle in the $U$ (internal energy) versus $\lambda$ (tuning parameter) diagram. The processes $A \rightarrow B$ and $C \rightarrow D$ are the quasistatic isothermal strokes performed at temperatures $T_h$ and $T_c$, respectively. The strokes $B \rightarrow C$ and $D \rightarrow A$ are two isochoric thermalization strokes performed at constant tuning parameters $\lambda_2$ and $\lambda_1$, respectively. Here, $Q_i$ ($i = 1, 2, 3, 4$) denotes the net heat exchange in each corresponding stroke of the cycle. The box under the cycle schematizes the regenerator. It stores the heat $|Q_2|$ released by the working substance during the $B \rightarrow C$ process and, during the $D \rightarrow A$ process, returns (all or part of) this stored heat to the system, thereby enabling heat regeneration and helping the cycle operate more efficiently. The arrow labeled $W_{\mathrm{cost}}$ indicates work input associated with this regeneration. Detailed explanations are given in the main text.}\label{fig1}
\end{figure}

As with the most extensively analyzed Otto and Carnot cycles, the Stirling cycle has also been generalized to the quantum domain in many studies~\cite{huang2014,gupta2021,lin2003,phung2025,xia2024,yin2018,he2002,lin2003b,lin2004,lin2005,huang2002,yin2020,yin2025,yin2017,chen2002,wu1998,zhao2017,paula2025,alcala2026,purkait2022,wang2024,xu2025,soufy2025,cruz2023,cakmak2023,das2023,castorene2025,castorene2025-2,hamedani2021,aydiner2021,rastegar2025}. In direct analogy with its classical counterpart, the quantum Stirling cycle consists of four strokes $-$ two quantum isochoric and two quantum isothermal processes $-$ as illustrated in Fig.~\ref{fig1}. The key point in the quantum construction is that, during all strokes of the cycle, the working substance is in thermal contact with  external thermal reservoirs~\cite{quan2007}. This situation is described within the standard open quantum systems framework, where the system-bath coupling is assumed to be weak and the Born-Markov and secular (rotating-wave) approximations are satisfied~\cite{dann2018,breuer2002}. Under these conditions, the dynamics of the working substance can be characterized by its reduced density matrix $\rho(t)$, which evolves according to a Lindblad-form quantum master equation, $\dot{\rho}(t) = -i/\hbar \big[ H(\lambda(t)), \rho(t) \big] + \mathcal{L}_t[\rho(t)]$. Here, $\hbar$ is the reduced Planck constant, $H(\lambda(t))$ is the Hamiltonian of the working substance, and $\lambda(t)$ denotes a time-dependent control (tuning) parameter that is externally driven. The dissipator $\mathcal{L}_t[\rho(t)]$ is a superoperator that encodes the irreversible thermal action of the reservoir on the system~\cite{dann2018,breuer2002}. Within this framework, the two quantum isothermal strokes are modeled as quasistatic processes, i.e., $\mathrm{d}\lambda(t)/\mathrm{d}t \rightarrow 0$, so that the state remains (approximately) Gibbsian at the bath temperature at each instant~\cite{quan2007}. This requires that the driving is much slower than the bath-induced relaxation. In contrast, during the isochoric strokes the control parameter $\lambda$ is kept fixed, and the role of the stroke is purely thermalization: by the end of the isochoric process (i.e., in the steady state), the working substance is assumed to have relaxed to a Gibbs (canonical) state at the temperature of the relevant heat bath~\cite{quan2007}. The processes of the standard four-stroke regenerative cycle, which assumes local thermal equilibrium at each point, are detailed below.

\textbf{(1) Isothermal process at $T_h$:} While the working substance, characterized by the Hamiltonian $H(\lambda)$, remains in contact with the hot bath at temperature $T_h$, the external control parameter $\lambda$ is slowly varied from $\lambda_1$ to $\lambda_2$ in a quasistatic manner. In this limit, the system is assumed to remain in thermal equilibrium with the hot bath at all times. The reduced density matrix at the initial and final points $i=A,B$ of this stroke is therefore given by $\rho_i = 1/Z_i \exp\left[-H(\lambda_i)/(k_B T_h)\right]$, where $k_B$ is the Boltzmann constant, $Z_i$ is the canonical partition function, and we identify $\lambda_A=\lambda_1$ and $\lambda_B=\lambda_2$. During this isothermal stroke, the net exchanged heat with the bath is denoted by $Q_1$.

\textbf{(2) Isochoric cooling from $T_h$ to $T_c$:} The working medium cools down to the temperature $T_c$ while being in contact with the regenerator. The tuning parameter is kept fixed at $\lambda_2$. The reduced density matrix of the working substance at point $C$ can then be expressed as a thermal state $\rho_C = 1/Z_C \exp\left[-H(\lambda_2)/(k_B T_c)\right]$. In the above description of the system-bath interaction, the regenerator is assumed to act as a macroscopic heat bath at temperature $T_c$, possessing an effectively infinite heat capacity and thus remaining at a constant temperature during the process. The system–bath interaction is weak, and the Born–Markov as well as the secular approximations are satisfied, ensuring that the system can indeed thermalize to the steady state. An amount of heat $Q_2$ is released by the working medium and stored in the regenerator.

\textbf{(3) Isothermal process at $T_c$:} While in contact with the cold bath at temperature $T_c$, the working substance undergoes a slow variation of the control parameter from $\lambda_2$ to $\lambda_1$, exchanging heat $Q_3$ with the bath while remaining in thermal equilibrium at $T_c$. The state at point $D$ is given by  $\rho_D = 1/Z_D \exp\left[-H(\lambda_1)/(k_B T_c)\right]$.

\textbf{(4) Isochoric heating from $T_c$ to $T_h$:} The temperature of the working substance increases to $T_h$ while it remains in thermal contact with the regenerator, with the tuning parameter kept fixed at $\lambda_1$. Consequently, the cycle closes at point $A$, described by the reduced thermal state $\rho_A$ as defined above. During this process, an amount of heat $Q_4$ is absorbed from the regenerator. In a similar manner, the regenerator is treated here as an effective Markovian bath at temperature $T_h$, since it thermalizes the working medium in the steady state.

The exchanged heat terms in the isothermal strokes (\textbf{(1)} and \textbf{(3)})  can be obtained using the relation $\bar{d} Q = T\,dS$, where $S$ denotes the von Neumann entropy~\cite{quan2007}. In the isochoric strokes (\textbf{(2)} and \textbf{(4)}), no work is performed; consequently, the change in the internal energy is entirely attributed to the heat exchange~\cite{quan2007}. The corresponding heat terms can be expressed as:
\begin{eqnarray}\label{heat}
Q_1 &=& T_h (S_B-S_A),\nonumber\\
Q_2 &=& U_C-U_B,\nonumber\\ 
Q_3 &=& T_c (S_D-S_C),\nonumber\\
Q_4 &=& U_A-U_D, 
\end{eqnarray}
where $S_i=-k_B \, \mathrm{tr}(\rho_i \ln \rho_i)$ and $U_i=\mathrm{tr}(\rho_i H(\lambda_i))$ refers to the internal energy. The net work for one cycle is determined using the first law as $W = \sum_i Q_i$.

Inspired by the classical Stirling cycle, the inclusion of a regenerator as a means to enhance the performance of the quantum cycle has been considered in several previous studies~\cite{huang2014,gupta2021,lin2003,phung2025,xia2024,yin2018,myers2022,he2002,lin2003b,lin2004,lin2005,huang2002,yin2020,yin2025,yin2017,chen2002,wu1998,zhao2017,paula2025,alcala2026}. Assuming a regenerator with no internal losses, its purpose is to store the heat released by the working medium during process \textbf{(2)} and to return it during process \textbf{(4)}. In other words, the regenerator functions to regenerate the heat $|Q_2|$. When an ideal-gas working medium undergoes the classical Stirling cycle~\cite{zohuri2018,cengel2019}, the heat $|Q_2|$ regenerated by the regenerator is fully sufficient to complete the cycle (that is, to return to point $A$ in Fig.~\ref{fig1}). Under this condition, the efficiency of the cycle can reach the classical Carnot efficiency. This situation, referred to as perfect regeneration, is enabled by the fact that the constant-volume heat capacity of an ideal gas is independent of temperature, together with the assumption of an ideal, lossless regenerator.

On the other hand, many studies have shown that perfect regeneration (i.e., $Q_4 = |Q_2|$) can only be achieved under certain specific conditions in quantum systems~\cite{huang2014,gupta2021,lin2003,phung2025,xia2024,yin2018,myers2022,he2002,lin2003b,lin2004,lin2005,alcala2026}. If the heat transfer between the working substance and the regenerator is defined as $\Delta Q = Q_2 + Q_4$, two distinct cases emerge. In the first case, $\Delta Q > 0$ (i.e., $Q_4 > |Q_2|$), the regenerator absorbs less heat during process \textbf{(2)} than it releases during process \textbf{(4)}. In this situation, the additional heat required to close the cycle must be supplied by the hot reservoir at temperature $T_h$. In the second case, $\Delta Q < 0$ (i.e., $Q_4 < |Q_2|$), the regenerator absorbs more heat during process \textbf{(2)} than it releases in process \textbf{(4)}. The excess heat remaining in the regenerator must then be released to the cold bath at temperature $T_c$ in order to prepare the regenerator for the next cycle. In both cases, the regenerator should not only recover and reuse heat, but also, when needed, act as a bridge for heat flow between the working substance and the external reservoirs — either providing extra heat from the hot bath or returning excess heat to the cold bath so that the cycle can run properly~\cite{huang2014,gupta2021,lin2003,phung2025,xia2024,yin2018,myers2022,he2002,lin2003b,lin2004,lin2005}.

Based on the regenerative behavior described above, the net heat transferred to the working substance from the hot side can be expressed as~\cite{huang2014,gupta2021,lin2003,phung2025,xia2024,yin2018,myers2022}
\begin{equation}\label{qin}
Q_h=Q_1+\max\{0,\Delta Q\}.    
\end{equation}
Here, the last term on the right-hand side of the equation contributes to the net absorbed heat when $\Delta Q > 0$, representing the additional heat that must be supplied to the working medium by the hot reservoir at $T_h$ in order to close the cycle. In a similar manner, the net heat rejected to the cold side is $Q_c=Q_3+\min\{0,\Delta Q\}$. The engine operates in heat-engine mode when $W=Q_h+Q_c>0$, which implies $Q_h>-Q_c>0$.

The framework of the quantum regenerative Stirling cycle discussed so far—defined by a four-stroke process where the working substance attains a Gibbs thermal state at the corresponding bath temperature at the end of each isochoric stroke—has recently been widely explored across various quantum working substances in the literature~\cite{huang2014,gupta2021,lin2003,phung2025,xia2024,yin2018,myers2022,he2002,lin2003b,lin2004,lin2005,huang2002,yin2020,yin2025,yin2017,chen2002,wu1998,zhao2017,paula2025,alcala2026}. In such studies, the efficiency has been defined in a standard way as the ratio of the net work output to the heat absorbed from the hot reservoir, i.e., $\eta = W / Q_h$. Refs.~\cite{huang2014,phung2025,zhao2017} have shown - and we will likewise demonstrate in the following analysis - that the efficiency can surpass the classical Carnot limit defined solely by the temperatures of the external reservoirs, $\eta_C = 1 - T_c/T_h$. It is worth noting here that this apparent performance gain beyond Carnot limits actually stems from the incomplete accounting of a mandatory work input; it relies on the regenerator, which, as we will show, implicitly requires an additional energy expenditure for the regeneration process—a cost not accounted for in the standard definition of $\eta = W / Q_h$~\cite{huang2014,myers2022}.

However, a fundamental challenge arises: determining the energy expended for regeneration without any knowledge of the regenerator's specific material structure or physical properties. Fortunately, by focusing on the thermodynamic task the regenerator performs, we can assign the minimum energy cost allowed by nature—the Carnot limit. Accordingly, our main claim is as follows. Based on the above construction of the cycle and on the open quantum system treatment of weak system-bath interactions, there exists an unavoidable hidden thermodynamic cost that cannot be neglected and should be included in the definition of efficiency. In the thermodynamic cycle analysis formulated in terms of the reduced dynamics of the working substance, the regenerator is effectively treated as a heat bath at temperature $T_c$ or $T_h$ during stroke \textbf{(2)} or \textbf{(4)}, respectively. In this formulation, it is implicitly implied that the regeneration process requires the heat $|Q_2|$ stored at the lower temperature level $T_c$ to be usable in a practical way at the higher temperature level $T_h$. This heat-upgrading process corresponds to pumping heat from a colder side to a hotter one, and by the second law of thermodynamics, it cannot occur without an external work input~\cite{cengel2019}; therefore, regeneration effectively entails an additional work requirement to make $|Q_2|$ usable at the higher temperature level $T_h$. In this work, we treat the required work input as a \emph{regeneration cost}, and we denote it by $W_{\mathrm{cost}}$.

The most conservative estimate for $W_{\mathrm{cost}}$ is obtained from the ideal Carnot refrigerator or heat-pump limit. Whether one considers extracting the heat $|Q_2|$ from the cold side at $T_c$ (refrigeration) or delivering the total conserved energy $|Q_2| + W_{\mathrm{cost}}$ to the hot side at $T_h$ (heat pumping), both perspectives necessitate the same minimum work input to make the heat $|Q_2|$ usable at the higher temperature level $T_h$, given by $W_{\mathrm{cost}} = |Q_2|((T_h - T_c)/T_c)$. Alternatively, this cost can be understood through the lens of exergy: the minimum energy required to upgrade an entropy unit, $\Delta S = |Q_2|/T_c$, stored at $T_c$ to the $T_h$ level is exactly $W_{\mathrm{cost}} = T_h \Delta S - T_c \Delta S = |Q_2|((T_h - T_c)/T_c)$. Furthermore, this $W_{\mathrm{cost}}$ perfectly restores the entropy balance during the regeneration process. If we were to transfer $|Q_2|$ directly from $T_c$ to $T_h$ without work, the entropy change of the effective reservoirs would be $\Delta S_{\mathrm{reg}} = -|Q_2|/T_c + |Q_2|/T_h < 0$, resulting in a direct violation of the second law. Including the reversible work input ensures the zero entropy production condition, $\Delta S_{\mathrm{reg}} = -|Q_2|/T_c + (|Q_2| + W_{\mathrm{cost}})/T_h = 0$. Fundamentally, these four thermodynamic approaches reveal that $W_{\mathrm{cost}}$ is an unavoidable energetic penalty, strictly required to make the heat $|Q_2|$ accessible to the working substance at the $T_h$ level during the $D \to A$ regeneration process. At the same time, all these fundamental thermodynamic approaches converge to a unique minimum cost for heat regeneration, given by:
\begin{equation}
W_{\mathrm{cost}} = |Q_2| \left( \frac{T_h - T_c}{T_c} \right).
\label{eq:Wcost}
\end{equation}
It should be noted that, without assigning a specific internal structure to the regenerator and basing the analysis solely on its thermodynamic function, the Carnot heat-pump limit given by Eq.~(\ref{eq:Wcost}) is the only unique, model-independent expression that can be formulated for this energy cost. Furthermore, more realistic regenerator models will inevitably involve internal irreversibilities and additional operational costs. Consequently, for such physical implementations, the actual regeneration cost will strictly exceed this ideal bound, requiring $W_{\mathrm{cost}} > |Q_2|((T_h - T_c)/T_c)$.

The final step is to incorporate the term $W_{\mathrm{cost}}$ into the cycle formulation. There is no general recipe for how this should be done~\cite{cangemi2024,gardas2015,thomas2018,guff2019,abah2017}. In quantum cycle analyses that involve system–quantum reservoir interactions, thermodynamic cost terms associated with creating or maintaining non-equilibrium states/correlations enter the thermodynamic equations in different ways, in order to show that the efficiency always remains bounded by the Carnot limit~\cite{cangemi2024,gardas2015,thomas2018}. Similarly, in studies of shortcuts to quantum adiabaticity, the so-called “cost of the shortcut” is typically introduced as a term that reduces the output work~\cite{guff2019} or as an additional energy input~\cite{abah2017}. Such \textit{ad hoc} incorporations usually depend on the specific role that the cost term is intended to play.

In standard thermodynamics, cycle efficiency is defined as the ratio of useful energy output to total energy input. Here, the regeneration cost, $W_{\mathrm{cost}}$, acts as external work supplied to make regeneration feasible at the $T_h$ level. When this cost is treated purely as an energetic input, it must be explicitly included in the total input energy~\cite{soufy2025,abah2017}. Accordingly, we define a modified efficiency for the regenerative quantum Stirling cycle as the net work produced per unit of total resource consumption, which can be expressed as:
\begin{equation}\label{efficiency}
\eta=\frac{W}{Q_h+W_{\mathrm{cost}}}.
\end{equation}
Here, $Q_h$ is given in Eq.~(\ref{qin}) and $W_{\mathrm{cost}} = |Q_2|\,((T_h - T_c)/T_c)$. Equation~(\ref{efficiency}) measures the global efficiency of the composite system, where the denominator represents the total energy bill paid. This total energetic cost encompasses the standard heat input from the hot reservoir, $Q_h$, alongside the additional energetic penalty, $W_{\mathrm{cost}}$, paid to operate the regenerative heat pump. It is also important to highlight that an alternative modified efficiency, defined as $\eta = (W - W_{\mathrm{cost}}) / Q_h$, is frequently used in the literature, particularly in shortcut-to-adiabaticity calculations~\cite{guff2019} to include the energetic cost of the shortcut—a protocol specifically designed to enhance the work and power output in finite-time operations. However, this definition is not appropriate for the present regenerative analysis. In our model, regeneration does not alter the net work; the system yields the exact same work output, $W$, in both regenerative and non-regenerative cycles. Furthermore, since the regenerator in our framework acts effectively as a thermal bath, the cycle duration—and consequently the power output—remains unchanged. Since regeneration is fundamentally a process aimed strictly at improving efficiency by recycling internal heat (thereby reducing the external heat demand) rather than boosting work or power, treating the energetic cost as an additional input in the denominator is the physically consistent approach. This modified efficiency definition constitutes a central result of the present study. In particular, it will be applied to two previously studied examples of regenerative Stirling cycles, which were reported to operate beyond the Carnot efficiency~\cite{huang2014}, in order to reassess their performance within a thermodynamically consistent framework.

As a working substance of a regenerative quantum Stirling cycle, we adopt two working media. One of them is a single spin-$1/2$ in a magnetic field, characterized by the Hamiltonian $H(\lambda)=1/2 \lambda \sigma_z$, and the other one is a pair of spin-$1/2$ particles in a magnetic field, interacting with each other through spin flip-flop processes and described by the Hamiltonian $H(\lambda)=1/2 \lambda (\sigma_z^1+\sigma_z^2)+J (\sigma_+^1 \sigma_-^2+\sigma_-^1 \sigma_+^2)$. Here, $\lambda$ is the tunable parameter in the cycle, $\sigma_z$ and $\sigma_{\pm}$ are the corresponding Pauli matrices, and $J$ denotes the antiferromagnetic coupling strength. In a recent paper~\cite{huang2014}, the regenerative quantum Stirling cycle was investigated for the two working media described above, and a high performance exceeding the Carnot limit was reported. Here, we revisit the quantum Stirling cycle by explicitly including the cost of regeneration. Here, we also analyze the conventional Stirling cycle without regeneration, in which the stages \textbf{(2)} and \textbf{(4)} are performed while the system is coupled to the standard cold and hot heat baths, respectively~\cite{alcala2026,purkait2022,wang2024,xu2025,soufy2025,cruz2023,cakmak2023,das2023,castorene2025,castorene2025-2,hamedani2021,aydiner2021,rastegar2025}. The efficiency in this case is given by $\eta=W/(Q_1+Q_4)$.

\begin{figure}[h]
\centering
\includegraphics[width=0.9\textwidth]{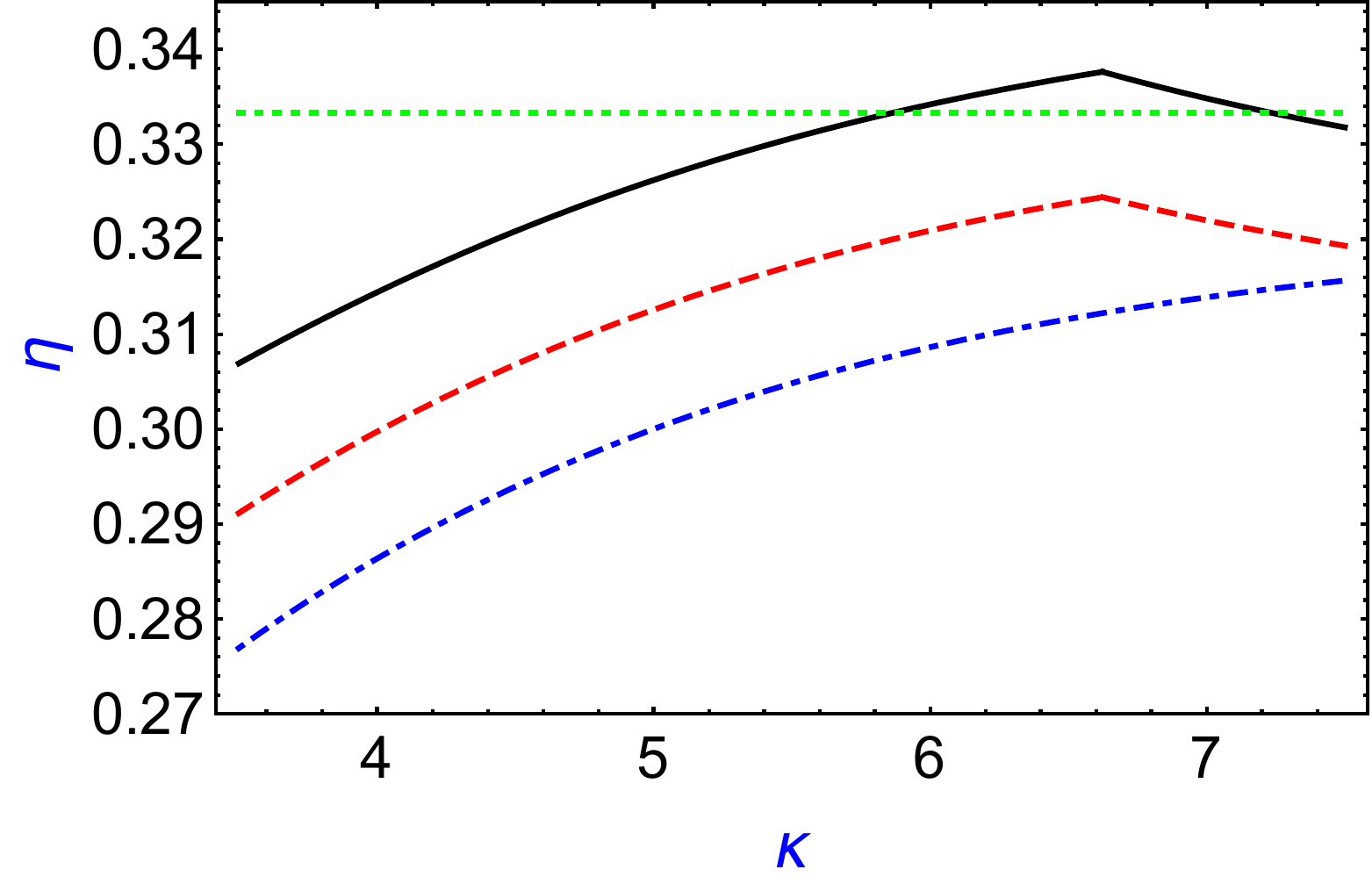}
\caption{Efficiency as a function of the relative magnetic field strength $\kappa$ ($\kappa=\lambda_1/\lambda_2$) for a single spin-$1/2$ working medium, with $\lambda_2=2.0$ and temperatures $T_h=3$ and $T_c=2$. The black solid curve shows the efficiency of the regenerative cycle without including any regeneration cost. The red dashed curve includes the regenerative cost $W_{\mathrm{cost}}$. The blue dot-dashed curve corresponds to the conventional Stirling cycle without a regeneration stage. The green dotted horizontal line marks the Carnot bound. We use dimensionless units where $\hbar=k_B=1$.}\label{fig2}
\end{figure}

\begin{figure}[h]
\centering
\includegraphics[width=0.9\textwidth]{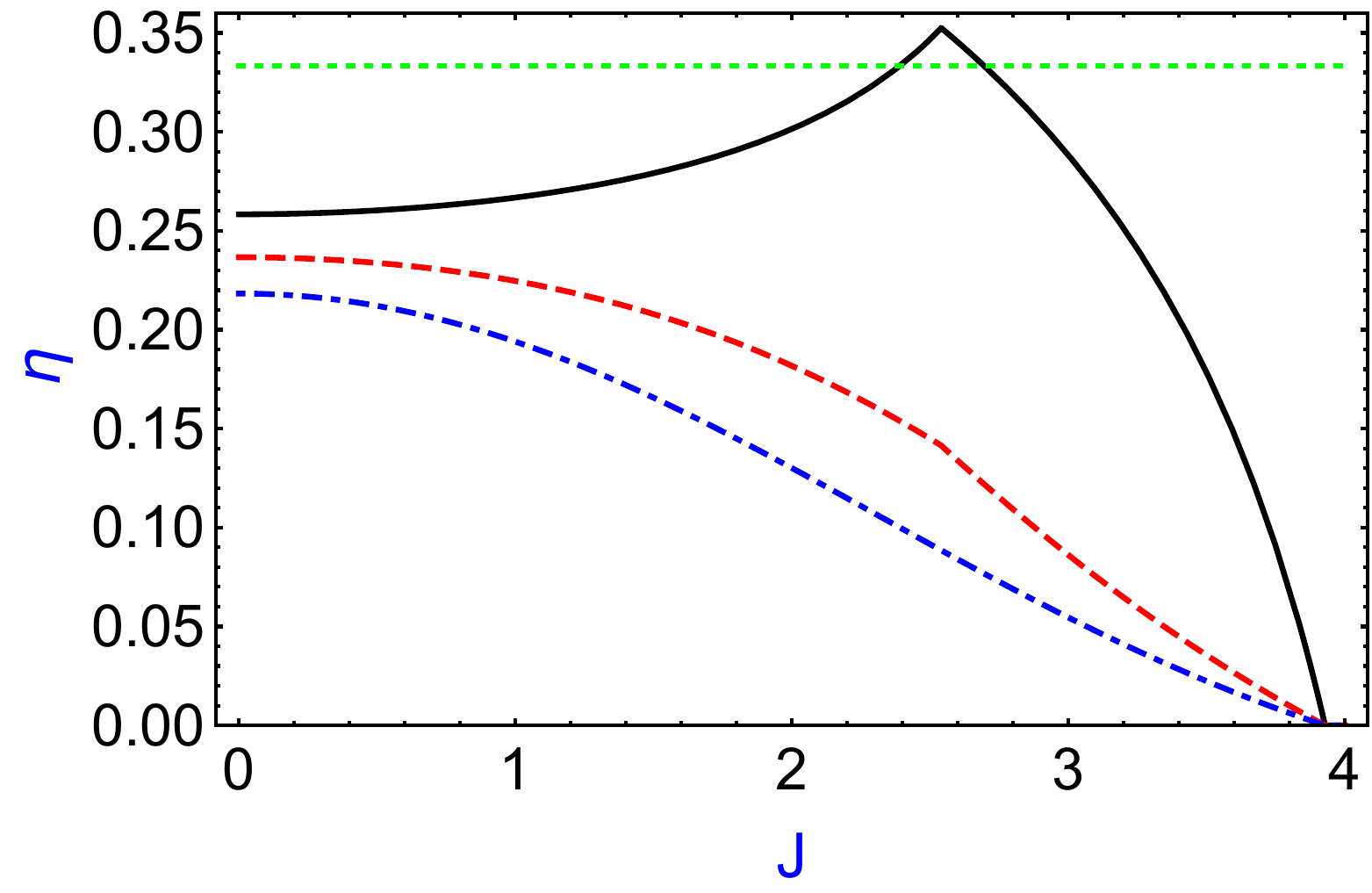}
\caption{Efficiency as a function of the interaction strength $J$ for a coupled-spins working medium, with temperatures $T_h=3$ and $T_c=2$, and tuning parameters $\lambda_1=2$ and $\lambda_2=1$. The line styles and color coding are the same as described in the caption of Fig.~\ref{fig2}.}\label{fig3}
\end{figure}

Fig.~\ref{fig2} depicts the efficiency as a function of the relative magnetic field strength $\kappa$ for a single spin-$1/2$ working medium, where $\lambda_1 = \kappa \lambda_2$, $\lambda_2 = 2.0$, $T_h = 3$, and $T_c = 2$. In a similar manner, Fig.~\ref{fig3} shows the case of two coupled spins, where the efficiency is plotted as a function of the coupling strength $J$ for $T_h = 3$, $T_c = 2$, $\lambda_1 = 2$, and $\lambda_2 = 1$. Qualitatively similar plots for the regenerative Stirling cycle, without the regenerative cost term ($W_\mathrm{cost}=0$), are also provided in Ref.~\cite{huang2014}. Here, we revisit those results by incorporating the thermodynamic cost of regeneration. We also present the corresponding results for the conventional Stirling cycle without regeneration, in order to clearly demonstrate the performance benefits introduced by the regenerative process. In both figures, the Carnot bound, $\eta_C = 1 - T_c/T_h$, is indicated by the horizontal dotted line. In both working-medium cases, the efficiency of the regenerative processes without accounting for any regeneration cost (solid lines) yields the highest performance, even surpassing the Carnot limit~\cite{huang2014,phung2025,zhao2017}. When the work input associated with regenerator preparation is included in the efficiency evaluation (dashed lines), the resulting efficiency is lower and remains below $\eta_C$; however, this not only provides numerical justification for the second law of thermodynamics but also demonstrates that a machine more efficient than the conventional Stirling cycle is still attainable. In Fig.~\ref{fig2}, the efficiency exhibits the same qualitative behavior as a function of the parameter $\kappa$ in the regenerative cases, with and without the regenerative cost included—showing an initial increase followed by a subsequent decrease—whereas the conventional Stirling cycle displays a purely monotonic increase. On the other hand, for the coupled-spin working medium (Fig.~\ref{fig3}), the system exhibits heat-engine behavior within the range $0 < J < 4$. In this regime, the efficiencies of both the conventional Stirling cycle and the regenerative Stirling cycle that includes the $W_{\mathrm{cost}}$ term show a monotonic decrease with increasing $J$. In contrast, the regenerative cycle efficiency evaluated without including the $W_{\mathrm{cost}}$ term displays a non-monotonic trend, characterized by an initial increase followed by a subsequent decrease as $J$ increases.

For the conventional Stirling cycle~\cite{alcala2026,purkait2022,wang2024,xu2025,soufy2025,cruz2023,cakmak2023,das2023,castorene2025,castorene2025-2,hamedani2021,aydiner2021,rastegar2025}, which involves no regeneration process and whose isochores are executed by coupling the working medium directly to the standard thermal reservoirs, the second law can be readily justified using calculation techniques similar to those developed, for example, in Refs.~\cite{altintas2025,campisi2016,peterson2019,camati2019,esposito2010,deffner2011}. Using the free-energy expression for Gibbs states, $F=U-TS=-k_B  T \ln{Z}$, the entropy at the instants of the isochores can be written as $S_i=U_i/T_i+k_B \ln{Z_i}$. The quantum relative entropy~\cite{nielsen2000}, $S(\rho_i||\rho_f)=tr\left(\rho_i\left(\ln{\rho_i}-\ln{\rho_f}\right)\right)$, provides the essential tool to justify our statement. For an isochore connecting two distinct thermal states of the same Hamiltonian $H(\lambda)$, namely $\rho_i=1/Z_i \exp{[-\beta_i H(\lambda)]}$ and $\rho_f=1/Z_f \exp{[-\beta_f H(\lambda)]}$, the relative entropy can be straightforwardly expanded as $S(\rho_i||\rho_f)=U_i\left(\beta_f-\beta_i\right)+\ln{\left(Z_f/Z_i\right)}$. Using the above tools, and after some straightforward algebra (see Appendix~\ref{app:Lag} for more details), the efficiency of the conventional Stirling cycle can be expressed as a deviation from the Carnot efficiency in terms of the relative entropies as
\begin{eqnarray}\label{lag1}
\eta=\eta_C-\frac{S(\rho_B||\rho_C)+S(\rho_D||\rho_A)}{\beta_c (Q_1+Q_4)}.
\end{eqnarray}
For the heat engine mode, $(Q_1 + Q_4) > 0$, and from the nonnegative property of the relative entropy~\cite{nielsen2000}, $S(\rho_i || \rho_f) \geq 0$, the second term on the right-hand side of Eq.~(\ref{lag1}) is always positive, which implies that $\eta < \eta_C$, thereby justifying the second law of thermodynamics for the conventional Stirling cycle. The deviation from the Carnot efficiency arises solely from the irreversibilities of the isochores, where the numerator accounts for the total (dimensionless) entropy production in the cycle as $\Sigma = S(\rho_B || \rho_C) + S(\rho_D || \rho_A) > 0$~\cite{altintas2025,campisi2016,peterson2019,camati2019,esposito2010,deffner2011} (see Appendix~\ref{app:Lag}).

For the regenerative Stirling cycle, incorporating the regeneration cost profoundly clarifies the thermodynamic landscape. If the regenerative work cost, $W_{\mathrm{cost}}$, is neglected, the efficiency artificially exceeds the Carnot limit, as previously reported in Refs.~\cite{huang2014,phung2025,zhao2017} and illustrated in Figs.~\ref{fig2} and~\ref{fig3}. However, when $W_{\mathrm{cost}}$ is rigorously included in the efficiency definition (Eq.~(\ref{efficiency})) and evaluated at its fundamental lower bound set by the Carnot heat-pump limit, the second law is strictly upheld. Yet, in contrast to the conventional Stirling cycle where the Carnot deficit ($\eta_{C} - \eta$) can be expressed in a manifestly positive closed form across all parameter regimes (as derived in Eq.~(\ref{lag1})), the derivation for the regenerative cycle is more intricate and does not yield such a universally positive analytical expression (see Appendix~\ref{app:Wcost}). Instead, the analytical proof depends explicitly on the operating temperature regimes. Crucially, our formalism allows us to analytically prove that the Carnot efficiency remains a definitive upper bound (i.e., $\eta_{C} - \eta > 0$) for all temperature regimes where $T_{h} \ge 2T_{c}$. In the regime where $T_{h} < 2T_{c}$, the Carnot deficit cannot be reduced to a strictly positive closed form. Nevertheless, the absence of a universally positive analytical proof in this specific parameter space does not signal a violation of the second law; rather, it reflects the inherent mathematical limitation of projecting a reversible macroscopic bound onto an irreversible, reduced quantum process. To decisively confirm the validity of the Carnot bound in this mathematically elusive $T_{h} < 2T_{c}$ regime, we point to our exact numerical results in Figs.~\ref{fig2} and~\ref{fig3}. For these evaluations, the bath temperatures are explicitly set to $T_{h}=3$ and $T_{c}=2$, perfectly capturing the $T_{h} < 2T_{c}$ regime. As depicted by the red dashed lines, the efficiency consistently and strictly remains below the Carnot limit, demonstrating the absolute physical robustness of our cost-inclusive approach across all explored parameter spaces.

The fundamental reason we cannot provide a universal analytical proof for $\eta_C - \eta \ge 0$ under this specific $W_{\mathrm{cost}}$ bound lies in the structural depth of the theoretical framework itself. The analysis of the hybrid working substance-regenerator system is conducted exclusively through the reduced state of the working substance, and the operational cost of the regenerator is introduced \textit{ad hoc} as a macroscopic penalty based on nature's absolute ideal limit---the Carnot heat-pump bound. Specifically, this macroscopic limit assumes zero internal entropy production in the regenerator. On the other hand, there is an inevitable, positive entropy production inherently originating from the irreversible isochoric thermalization of the working substance; therefore, the Carnot efficiency naturally stands as the strict upper limit. Nature inherently protects its fundamental laws: the physical reality of the second law dictates that the actual regenerative mechanism must strictly demand a higher energetic cost ($W_{\mathrm{cost}}$) than the ideal, reversible Carnot limit precisely to compensate for this localized entropy production. This approach aligns perfectly with established methodologies in the literature; for instance, in hybrid quantum Otto cycle analyses involving system-quantum reservoir interactions, thermodynamic consistency is guaranteed by imposing an exact working substance parameter-dependent penalty term on the heat exchanges to enforce the Carnot bound~\cite{cangemi2024,gardas2015,thomas2018}. Following this exact logic, instead of attempting to force a general proof from the purely macroscopic Carnot limit, we use Eq.~(\ref{Wbound2}) in Appendix~\ref{app:Wcost} to provide a sufficient, parameter-dependent lower bound on $W_{\mathrm{cost}}$ that mathematically guarantees $\eta_C - \eta \ge 0$ across all regimes. Ultimately, the second law imposes a rigorous master condition, dictating that the true regeneration cost must satisfy the maximum of both the macroscopic thermodynamic constraint and the microscopic bound (see Appendix~\ref{app:Wcost} for the full derivation details):
\begin{equation}\label{Wbound}
W_{\mathrm{cost}} \geq \max\left\{|Q_2|(T_h - T_c)/T_c,\left(Q_1+Q_4-Q_h\right)-\frac{\Sigma}{\beta_c \eta_C}\right\}.
\end{equation}

As discussed, the current analytical bottleneck in universally proving the Carnot limit arises from treating the regenerator phenomenologically as a macroscopic reservoir. Moving beyond this hybrid approach and modeling the regenerator as an active quantum component will naturally resolve this mathematical conflict, allowing the Carnot bound to be analytically verified across all regimes. Therefore, to achieve a fully consistent and physically realistic framework in future studies, the system-regenerator coupling and correlations must be explicitly included. To this end, three candidate quantum-regenerator models suggest themselves: First, the regenerator could be modeled not as a Markovian heat bath but as a non-Markovian reservoir~\cite{thomas2018,breuer2009,guarnieri2016}. Unlike memoryless Markovian baths, non-Markovian reservoirs do not necessarily cause irreversible information loss and can exhibit information and energy back-flow~\cite{guarnieri2016}, allowing a portion of the heat absorbed at stage \textbf{(2)} to be returned. Second, the regenerator can be modeled as an auxiliary quantum system on the same footing as the working substance~\cite{quan2007,peterson2019,rodriguez2016,alhambra2019,levy2016,kose2019}. The heat rejected during the stroke \textbf{(2)} can induce eigenstate transitions in this auxiliary system, thereby storing energy in its populations. Algorithmic or adiabatic control protocols can subsequently be applied to reconvert the stored energy and regenerate the heat during the stage \textbf{(4)}. Third, the regenerator may be described within a repeated–interaction (collision–model) framework, in which the working medium successively interacts with a finite chain of ancillary quantum systems that play the role of a structured reservoir with memory~\cite{ciccarello2022,lorenzo2017,scully2003,dillen2009}. During stage \textbf{(2)}, a subset of these ancillas absorbs energy from the working medium, while during stage \textbf{(4)} the same ancillas are brought back into contact with the working medium and partially release the stored energy, thereby allowing partial heat regeneration. In all of the above proposals, the quantum modeling of the regenerator is intrinsically tied to the specific structure of the working substance; therefore, it is generally not possible to formulate a universal, model-independent quantum-regenerator description. Moreover, the isochoric strokes can terminate in non-equilibrium states. Consequently, additional thermalization steps are required prior to the subsequent isothermal contacts, so that the full cycle can be naturally described as a modified six-stroke regenerative Stirling cycle. Given that formulating a fully quantum regenerator model within any of these frameworks requires developing a dedicated, complex open-system architecture—each of which constitutes a standalone research problem in its own right—a rigorous mathematical and thermodynamic investigation of these possibilities is well beyond the scope of the present study and is therefore left for future work.

\section{Conclusion}
We investigated the standard four-stroke regenerative quantum Stirling cycle, which assumes local thermal equilibrium at each stage, within the standard weak-coupling, Markovian open quantum system framework. Inspired by the classical regenerator model, to date, virtually all treatments model the regenerator as a passive heat buffer that stores heat during the isochoric stroke \textbf{(2)} and returns it in stroke \textbf{(4)}, with the corresponding work cost typically left implicit. In the reduced open-system description, the regenerator is treated as an effective heat reservoir that brings the working substance to thermal equilibrium at the end of the isochoric strokes. When this idealized, cost-free regenerator picture is adopted, the resulting bookkeeping may suggest very high apparent efficiencies, in some cases even exceeding the Carnot bound. In this work, based on the weak-coupling open quantum system description of the system-bath interactions, we point out that the operation of the regenerator implicitly requires the heat stored at a low temperature level to be made usable at a higher temperature level, i.e., that the stored (recovered) heat be pumped from the lower temperature to the higher temperature. By the second law of thermodynamics, this process necessarily requires an additional work input. We then accounted for this hidden work input as an explicit regeneration cost within the working-substance thermodynamic description, and we used it to redefine the cycle efficiency by taking the net work produced by the working substance per unit total energy input. We revisited the regenerative Stirling cycle using two working substances: a single spin-$1/2$ and a pair of interacting spin-$1/2$ particles. In both cases, we re-evaluated the cycle performance using our modified efficiency, which includes the regeneration cost taken at its minimum value set by the Carnot heat pump limit. For comparison, we also analyzed the conventional Stirling cycle without regeneration and computed its efficiency under the same conditions, in order to assess whether the regenerator still provides an advantage despite this cost. Our results demonstrate that the modified efficiency including $W_{\mathrm{cost}}$ remains below the Carnot efficiency, while still being higher than the efficiency of the conventional Stirling cycle. For the conventional Stirling cycle, we proved that the Carnot efficiency is a rigorous upper bound by using the quantum relative entropy formulation. In contrast, for the regenerative Stirling cycle within our phenomenological reduced description, we showed that the Carnot efficiency can be guaranteed as an upper bound provided that the regeneration cost $W_{\mathrm{cost}}$ exceeds a sufficient lower bound. Finally, to motivate future work, we proposed three possible quantum models in which the regenerator is treated as an active component of the cycle, and we briefly discussed the physical motivation behind each proposal.

\begin{itemize}
\item \textbf{Data Availability Statement} Data will be made available upon reasonable request.
\end{itemize}

\backmatter

\begin{appendices}

\section{Proof of Eq.~(\ref{lag1}) and entropy production}\label{app:Lag}
At the thermal points $(i=A,B,C,D)$, where the density matrix of the working medium is in Gibbs form  $\rho_i = Z_i^{-1} \exp\left[-\beta_i H(\lambda_i)\right]$, the free energy can be written as
\begin{equation}
F_i=U_i-T_i S_i=-k_B T_i \ln{Z_i}.
\end{equation}
Using this relation, the thermodynamic entropy can be written as
\begin{equation}\label{entropy}
S_i=k_B \beta_i U_i+k_B \ln{Z_i}.
\end{equation}
Now we expand the relative entropy $S(\rho_i||\rho_f)$ for the isochoric strokes that connect two different Gibbs states, $\rho_i=Z_i^{-1} \exp{[-\beta_i H(\lambda)]}$ and $\rho_f=Z_f^{-1} \exp{[-\beta_f H(\lambda)]}$, which share the same Hamiltonian $H(\lambda)$ but correspond to different temperatures. Using the logarithmic identity for a thermal (Gibbs) state, $\ln\{Z^{-1}\exp\left[-\beta H(\lambda)\right]\}=-\beta H(\lambda)-\ln{Z}$, together with the definition of the internal energy  $U=\mathrm{tr}(\rho H(\lambda))$ and the normalization $\mathrm{tr}(\rho)=1$, the relative entropy can be written as
\begin{eqnarray}\label{relativeentropy}
S(\rho_i||\rho_f)&=&tr(\rho_i\ln{\rho_i})-tr(\rho_i\ln{\rho_f})\nonumber\\
&=&U_i\left(\beta_f-\beta_i\right)+\ln{\left(Z_f/Z_i\right)}.  
\end{eqnarray}
Now we relate the efficiency deficit, $\eta_C-\eta$, to the corresponding relative entropies, as given in Eq.~(\ref{lag1}). Using the heat definitions in Eq.~(\ref{heat}), together with the entropy expression in Eq.~(\ref{entropy}) and the relative-entropy result in Eq.~(\ref{relativeentropy}), the identity in Eq.~(\ref{lag1}) follows straightforwardly. The key intermediate results required to arrive at Eq.~(\ref{lag1}) are given below as:
\begin{eqnarray}\label{lag2}
\eta_C-\eta&=&\frac{-T_c \left(Q_1+Q_4\right)-T_h \left(Q_2+Q_3\right)}{T_h\left(Q_1+Q_4\right)}\nonumber\\
&=&\frac{1}{\beta_c\left(Q_1+Q_4\right)}\left\{U_B(\beta_c-\beta_h)+\ln{\left(Z_C/Z_B\right)}+U_D(\beta_h-\beta_c)+\ln{\left(Z_A/Z_D\right)}\right\}\nonumber\\
&=&\frac{S(\rho_B||\rho_C)+S(\rho_D||\rho_A)}{\beta_c (Q_1+Q_4)}.
\end{eqnarray}

Now we focus on the total entropy production over the cycle. Since the isothermal processes are quasistatic, they produce no entropy. Along the isochoric strokes, the total entropy production has two contributions: one from the working substance and the other from the heat bath. Using the entropy expression in Eq.~(\ref{entropy}) and the relative-entropy expansion in Eq.~(\ref{relativeentropy}), for the stroke $B \rightarrow C$ the total entropy production $\Sigma_{B \rightarrow C}$ can be directly related to the relative entropy, thereby establishing its non-negative character:
\begin{eqnarray}
\Sigma_{B \rightarrow C} &=& (S_C-S_B)-\frac{1}{T_c} \left(U_C-U_B\right)\nonumber\\
&=&k_B\left\{U_B (\beta_c-\beta_h)+\ln{\left(Z_C/Z_B\right)}\right\}\nonumber\\
&=&k_B S(\rho_B||\rho_C).
\end{eqnarray}
Using similar steps for the stroke $D \rightarrow A$, one can show that
\begin{eqnarray}
\Sigma_{D \rightarrow A}&=& (S_A-S_D)-\frac{1}{T_h} \left(U_A-U_D\right) \nonumber\\
&=&k_B S(\rho_D\|\rho_A).
\end{eqnarray}
The (dimensionless) entropy production over the cycle can then be defined as $\Sigma=\left(\Sigma_{B \rightarrow C}+\Sigma_{D \rightarrow A} \right)/k_B=S(\rho_B||\rho_C)+S(\rho_D\|\rho_A)>0$. Based on this result and Eq.~(\ref{lag2}), the efficiency reduction due to the net entropy production takes the generic form~\cite{campisi2016,peterson2019,camati2019}: $\eta=\eta_C-\Sigma/(\beta_c Q_{\mathrm{in}})$, where $Q_{\mathrm{in}}$ denotes the net heat absorbed from the source.

\section{Proof of Eq.~(\ref{Wbound}) }\label{app:Wcost}
Here, we provide the details of the calculation leading to the inequality in Eq.~(\ref{Wbound}). Since both the conventional and the regenerative Stirling cycles connect the same states $A-B-C-D$ and no work is performed on the isochores, their net work outputs are equal. Using Eq.~(\ref{lag1}), the work output can be written in terms of relative entropies as follows:
\begin{eqnarray}\label{work-ent}
W=\eta_C \left(Q_1+Q_4\right)-\frac{\Sigma}{\beta_c}. 
\end{eqnarray}
We now calculate the "Carnot deficit", $\eta_C - \eta$, for the regenerative Stirling cycle using the modified efficiency in Eq.~(\ref{efficiency}) and the work output defined in Eq.~(\ref{work-ent}), as follows:
\begin{eqnarray}\label{lag3}
\eta_C-\eta&=&\left(1-\frac{\beta_h}{\beta_c}\right)-\frac{\beta_c\eta_C(Q_1+Q_4)-\Sigma}{\beta_c(Q_h+W_{\mathrm{cost}})}\nonumber\\
&=&\frac{\Sigma}{\beta_c(Q_h+W_{\mathrm{cost}})}+\frac{\beta_c\eta_C\left(Q_h+W_{\mathrm{cost}}-(Q_1+Q_4)\right)}{\beta_c(Q_h+W_{\mathrm{cost}})}.
\end{eqnarray}
Eq.~(\ref{lag3}) is the simplest expression that can be given for the Carnot deficit of the regenerative cycle. The first term on the right-hand side is always non-negative. On the other hand, under general conditions, the sign of the second term and its exact relationship with the first term remain analytically uncertain. When we evaluate this expression by setting $W_{\mathrm{cost}}$ to its theoretical lower bound determined by the Carnot heat pump limit, i.e., $W_{\mathrm{cost}} = |Q_2|\,((T_h - T_c)/T_c)$, the sign of the second term is unconditionally non-negative for temperature regimes where $T_h \ge 2T_c$. In this regime, we therefore always obtain $\eta_C > \eta$.  However, in the regime where $T_h < 2T_c$, this second term can become negative. Crucially, a negative second term does not imply that the Carnot deficit ($\eta_C - \eta$) becomes negative. The first term in the equation remains positive and generally dominates the second term, preserving the condition $\eta_C - \eta > 0$. This behavior is explicitly demonstrated by the exact numerical results presented in Figs.~\ref{fig2} and~\ref{fig3}. In these plots, the bath temperatures are set to $T_h = 3$ and $T_c = 2$, which precisely falls into the $T_h < 2T_c$ regime. As seen in the figures, even in this analytically ambiguous regime, the evaluated efficiency consistently stays strictly below the Carnot limit (see red dashed lines). Consequently, the absence of a closed-form analytical proof in this regime does not signal a violation of the second law, but rather reflects the mathematical limitation of applying a reversible macroscopic bound to an irreversible reduced quantum process. Instead of attempting to force a general proof from the Carnot limit, we use Eq.~(\ref{lag3}) to provide a sufficient lower bound on $W_{\mathrm{cost}}$ that mathematically guarantees $\eta_C - \eta \ge 0$ across all regimes. Since $(Q_h + W_{\mathrm{cost}}) > 0$, we obtain a sufficient lower bound on $W_{\mathrm{cost}}$ ensuring that the numerator of Eq.~(\ref{lag3}) is non-negative:
\begin{equation}\label{Wbound2}
W_{\mathrm{cost}} \geq \left(Q_1+Q_4-Q_h\right)-\frac{\Sigma}{\beta_c \eta_C}.
\end{equation}
In addition to Eq.~(\ref{Wbound2}), the second law imposes another condition, namely
$W_{\mathrm{cost}} \ge |Q_2|(T_h - T_c)/T_c$. Thus, the regeneration cost must satisfy both the thermodynamic constraint imposed by the Carnot limit and the lower bound given in Eq.~(\ref{Wbound2}). In other words, we obtain the stricter bound:
\begin{equation}\label{Wbound3}
W_{\mathrm{cost}} \geq \max\left\{|Q_2|(T_h - T_c)/T_c,\left(Q_1+Q_4-Q_h\right)-\frac{\Sigma}{\beta_c \eta_C}\right\}.
\end{equation}
\end{appendices}

\end{document}